\documentclass[twocolumn,showpacs]{revtex4}
\usepackage{amsmath}
\usepackage{mathptmx}
\usepackage{bm}
\usepackage{color}
\usepackage[english]{babel}
\usepackage{graphicx}

\begin{document}

\title{Semi-relativistic effects in spin-1/2 quantum plasmas}
\author{Felipe A. Asenjo}
\altaffiliation[Current address: ]{Institute for Fusion Studies, University of Texas, Austin, USA}
\affiliation{Department of Physics, Ume{\aa } University, SE--901 87 Ume{\aa }, Sweden}
\author{Jens Zamanian}
\affiliation{Department of Physics, Ume{\aa } University, SE--901 87 Ume{\aa }, Sweden}
\author{Mattias Marklund}
\affiliation{Department of Physics, Ume{\aa } University, SE--901 87 Ume{\aa }, Sweden}
\author{Gert Brodin}
\affiliation{Department of Physics, Ume{\aa } University, SE--901 87 Ume{\aa }, Sweden}
\author{Petter Johansson}
\affiliation{Department of Physics, Ume{\aa } University, SE--901 87 Ume{\aa }, Sweden}

\begin{abstract}
Emerging possibilities for creating and studying novel plasma regimes, e.g.
relativistic plasmas and dense systems, in a controlled laboratory
environment also requires new modeling tools for such systems. This brings
motivation for theoretical studies of the kinetic theory governing the
dynamics of plasmas for which both relativistic and quantum effects occur
simultaneously. Here, we investigate relativistic corrections to the Pauli
Hamiltonian in the context of a scalar kinetic theory for spin-$1/2$ quantum
plasmas. In particular, we formulate a quantum kinetic theory that takes
such effects as spin-orbit coupling and Zitterbewegung into account for the
collective motion of electrons. We discuss the implications and possible
applications of our findings.
\end{abstract}
\maketitle

\section{Introduction}

Plasmas, in their full generality, make up a highly complex class of
physical systems, from classical tenuous plasmas, in e.g.\ fluorescent
lighting, to dense, strongly coupled systems, such as QCD plasmas. The large
span of plasma systems implies that a wide variety of theoretical methods
have been developed for their treatment. Even so, there are general
principles that remain as common features between the different plasma
systems. Therefore, methods used for one plasma system can in some cases be
transferred to another plasma type, sometimes leading to new insights. One
such example is the transferal of techniques for treating nonlinearities in
classical plasmas to quantum mechanical plasmas. The latter, often termed
quantum plasmas (see, e.g., Refs.\ \cite%
{Pines,Manfredi,Melrose,Shukla-Eliasson}), to lowest order contains
corrections due to the classical regime in terms of nonlocal terms, related
to the tunneling aspects of the electron (in quantum plasmas, the ions are
most often treated classically). Such tunneling effects can be incorporated
in both kinetic and fluid descriptions of the collective electron motion 
\cite{Markowich-etal,Haas-etal}. Such collective tunneling effects may,
e.g., lead to nanoscale limitations in plasmonic devices \cite{Marklund-etal}
and bound states near moving test charges in plasmas \cite{Else-etal}.
Another mean-field effect that may be added to the dynamics of classical
plasmas concerns the electron spin, i.e. the possibility of large-scale
magnetization of plasmas \cite{Marklund-Brodin}. Such magnetization
switching is known to be able to give new non-trivial features, such as
metamaterial properties, allowing for, e.g., new soliton modes \cite%
{Pendry,Brodin-Marklund}. Moreover, the inclusion of the electron spin into
the collective dynamics can be done either through a fluid or a kinetic
approach \cite{Zamanian-etal}. Furthermore, the inclusion of spin into the
dynamics of a quantum plasma points in the direction of relativistic effects
in such plasma systems, e.g. collective spin-orbit coupling. It is the
intention of the present work to extend previous work into the weakly
relativistic regime.

As indicated above, the dynamics of plasmas under extreme conditions is an
important and integral part at many current and up-coming experimental
facilities, and such investigations therefore constitutes a highly active
research field. In particular, laboratory plasmas, such as laser generated
plasmas, are currently presenting the possibility of studying previously
unattainable plasma density regimes. It is well-known that in the
high-density regime \cite{glenzer-redmer} quantum effects start to play a
role for, e.g., the dispersive properties of plasma waves \cite%
{neumayer,fustlin,glenzer}. In nature, such dense relativistic plasmas can
be found in planetary interiors and in stars \cite{Ichimaru}. Moreover,
relativistic contributions to such plasma dynamics are under many
circumstances very important \cite{ross}. Thus, when the parameters takes
values characteristic for the quantum relativistic regime, one needs to
consider more complex dynamical models in order to obtain accurate
descriptions of a host of phenomena. A canonical starting point for dealing
with high-density effects in a perturbative relativistic regime is offered
by the quantum kinetic approach, here based on the Dirac description.
Effects that can be included in such a perturbative model includes, e.g.,
spin dynamics, spin-orbit coupling, and Zitterbewegung. These examples have
close connections to the nonperturbative relativistic quantum regime, in
which e.g. pair production \cite%
{Nerush-etal,Hebenstreit-etal1,Hebenstreit-etal2,Elkina-etal} and other
nonlinear quantum vacuum effects \cite{Marklund-Shukla} become pronounced

\begin{figure}[t]
\begin{center}
\includegraphics[scale = 0.24]{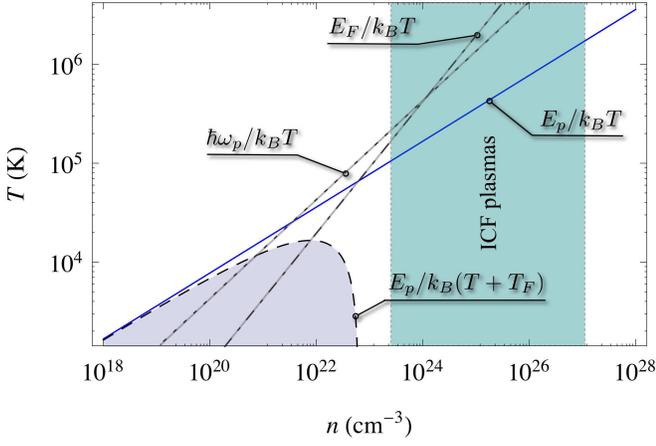}
\end{center}
\caption{Different plasma regimes in the temperature-density parameter
space. The dotted line is where the strong coupling parameter $\Gamma =
E_p/k_B T = 1$, where $E_p $ is the potential energy due to the nearest
neighbor. For higher densities the average kinetic energy of the particles
is given by the Fermi energy $k_B T_F$ rather than the thermal energy and
the strong coupling parameter must there be replaced by $\Gamma_F = E_p /
k_B (T + T_F)$. In this graph $\Gamma_F = 1$ is illustrated by the dashed
curve, and the strong coupling region (shaded) lies below this line. In this
region higher order correlations has to be taken into account, and the
mean-field description is not valid. For comparison, the lines $\hbar 
\protect\omega_p / k_B T$ (dotted gray line), where $\hbar \protect\omega_p$
is the plasmon energy, and $T_F /T$ (dotted-dashed gray line) are also
plotted. These measure, respectively, the relative importance of wave
function dispersion and the Fermi pressure. As a rough estimate, the quantum
regime lies below these lines. The area marked ICF denote the regime of
relevance for inertial confinement fusion experiments.}
\label{fig:regimes}
\end{figure}

Dense plasmas, and in particular short-time scale phenomena therein, have
been successfully studied using Green's functions techniques, such as the
Kadanoff-Baym kinetic equations \cite{Kadanoff-Baym} (see also Refs.\ \cite%
{Zwanzig,Prigogine,Barwinkel,Klimontovich} for similar approaches, and Ref.\ 
\cite{Kremp-etal} for an overview and examples from femtosecond laser
physics). Although the Kadanoff-Baym equations and similar indeed gives
ample opportunity to treat a wide variety of systems, their generality also
makes simplifying assumptions necessary, and under certain circumstances a
mean-field model, that still retains memory effects and non-local
structures, can be an adequate approximation \cite{Ichimaru}. In particular
the mean-field approach is well suited outside the regime of strong coupling
effects. Here, we will be interested in phenomena in plasmas that are not
strongly coupled, but still in regimes where a classical plasma descriptions
is not fully adequate. Here we stress that a large number of different
dimensionless parameters (see e.g. Refs. \cite{BMM-2008,Lundin2010} for more
complete discussions) is needed to give a thorough description of \ various
plasma regimes. However, much insight can be gained by considering a simple
density-temperature plot. In Figure \ref{fig:regimes} a schematic view of
the parameter regime of interest for our study is presented, adopted from
Ref. \cite{zamanian}

In Section \ref{1} we recall the Foldy-Wouthuysen transformation for
particles in external fields and as a result we obtain a semi-relativistic
Hamiltonian. We then go on to define the scalar quasi-distribution function
for spin-1/2 particles in Section \ref{2}. In Section IV we then derive the
corresponding evolution equation for the distribution function. The
evolution equation clearly depicts the importance of the different terms of
the relativistic expansion. A comparison of our results to previous studies
is made, and we discuss the applicability of our equation as well as the
interpretation of the variables involved. In Section V our theory is
illustrated by means of two examples from linearized theory, and finally, in
Section VI, our main conclusions are summarized.

\section{The high-order corrections}

\label{1}

The Dirac Hamiltonian can be written in the form 
\begin{equation}  \label{eq:form}
\hat{H} = \hat\beta mc^2 + \Omega_e + \Omega_o\, ,
\end{equation}
where we have the \textit{even} ($\hat \beta \Omega_e = \Omega_e \hat \beta$%
) operator 
\begin{equation}
\Omega_e = q\phi \, ,
\end{equation}
and the \textit{odd} ($\hat \beta \Omega_o = - \Omega_o \hat \beta $)
operator 
\begin{equation}
\Omega_o = c\bm{\hat\alpha} \cdot \left( \hat{\mathbf{p}} - q\mathbf{A}
\right)\, ,
\end{equation}
where $\hat\beta$ and $\bm{\hat\alpha}$ are the $4 \times 4$ Dirac matrices, 
$m$ is the electron mass, $q$ is the charge ($q=-e$ for an electron), $c$ is
the speed of light, $\hat{\mathbf{p}}$ is the momentum operator and $\phi$
and $\mathbf{A}$ are the scalar and vector potential, respectively.

The odd operators in the Dirac Hamiltonian couples the positive and negative
energy states of the Dirac bi-spinor. For the purpose of obtaining a
perturbative expansion in the parameter $E/mc^2$, where $E$ is the typical
energy associated with the second and third term in (\ref{eq:form}), we
assume that the first term in (\ref{eq:form}) is large compared to these
terms. The consecutive application of the unitary Foldy--Wouthuysen (FW) 
\cite{foldy-wouthuysen} transformation 
\begin{equation}
\hat{H} \rightarrow \exp\left( \beta\Omega_o/2mc^2 \right) \hat{H}
\exp\left( -\beta\Omega_o/2mc^2 \right)\, ,
\end{equation}
yields a new Hamiltonian of the form (\ref{eq:form}) in which the new odd
operators are of the order $1/mc^2$. Performing this transformation $n$
times yields terms up to order $(mc^2)^{-n}$. This gives a separation of the
positive and negative energy states up to an arbitrary order $n$ in $1/mc^2$.

Applying this transformation four times gives the following Hamiltonian for
positive energy states with only even operators \cite{strange} 
\begin{align}
\hat{H}& =mc^{2}+q\phi +\frac{1}{2m}\left( \hat{\mathbf{p}}-\frac{q}{c}%
\mathbf{A}\right) ^{2}-\frac{q\hbar }{2mc}\bm{\sigma}\cdot \mathbf{B}+\frac{%
\hbar ^{2}q}{8m^{2}c^{2}}\nabla \cdot \mathbf{E}  \notag \\
& -\frac{\hbar q}{4m^{2}c^{2}}\bm{\sigma}\cdot \left[ \mathbf{E}\times
\left( \hat{\mathbf{p}}-\frac{q}{c}\mathbf{A}\right) \right] -\frac{i\hbar
^{2}q}{8m^{2}c^{2}}\bm{\sigma}\cdot \nabla \times \mathbf{E}  \notag \\
& +\frac{1}{8m^{3}c^{2}}\left( \hat{\mathbf{p}}-\frac{q}{c}\mathbf{A}\right)
^{4}  \label{eq:pos}
\end{align}%
%
where $\bm{\sigma}$ here denotes a vector containing the $2\times 2$ Pauli
matrices, $\hbar $ is the reduced Planck constant, $q$ is the charge, $m$ is
the mass, $\mathbf{B}$ and $\mathbf{E}$ are the magnetic and electric field
and $\phi $ and $\mathbf{A}$ are the corresponding potentials in the Coulomb
gauge. We see that the first four terms constitute the Pauli Hamiltonian,
while the remaining terms are higher order corrections. In particular, the
sixth and seventh terms together gives Thomas precession and spin-orbit
coupling, while the fifth and eight terms are the Darwin term and the so
called mass-velocity correction term, respectively.

In Eq. (\ref{eq:pos}) as well as in the Dirac theory we started from, the
value of the spin $g$-factor is exactly 2. When applying the resulting
theory, in section V, we will use the QED corrected value of $g\simeq
2.00232 $, however. In spite of the smallness of the modification it turns
out that this correction is important, as the applications of our theory are
very sensitive to the exact value of the $g$-factor. In fact, the
sensitivity of the kinetic theory to the value of $g$ was seen already in
Ref. \cite{brodin}. This may suggest that for consistency, the Hamiltonian
for QED-corrections should be added to Eq. (\ref{eq:pos}). Such an approach
would indeed modify the $g$-value to the desired one in the theory presented
below, but the augmented Hamiltonian would also add several new terms in the
evolution equation for the electrons. Those extra terms are at least smaller
than those kept by a factor of the order $(g-2)$, however. Thus the main
effect from QED in the regime of study is the modification of the value of
the $g$-factor as compared to the Dirac theory. As a consequence, the
contributions from QED (see e.g. Ref. \cite{Groot-book} for QED-corrections
to the Dirac Hamiltonian) besides modifying the $g$-value will not be
included here.

\section{Gauge-invariant Stratonovich-Wigner function}

\label{2}

The extended phase-space scalar kinetic model is obtained using the
Hamiltonian \eqref{eq:pos}. Following Ref.~\cite{zamanian}, we are able to
construct a gauge invariant scalar kinetic theory using a density matrix
description for a spin-1/2 particle.

The basis states are $|\mathbf{x},\alpha\rangle=|\mathbf{x}\rangle\otimes
|\alpha\rangle$, where $|\mathbf{x}\rangle$ is a state with position $%
\mathbf{x}$ and $|\alpha\rangle$ is the state with spin-up $\alpha=1$ or
spin-down $\alpha=2$. As a starting point of this model we use the spinor
state $\psi\left( \mathbf{x},\alpha,t\right)=\langle\mathbf{x}%
,\alpha|\psi\rangle$ which fulfill the dynamical equation $i\hbar \partial_t
\psi\left( \mathbf{x},\alpha,t\right)=\hat H\psi\left( \mathbf{x},\alpha, t
\right)$, with the Hamiltonian \eqref{eq:pos}.

With the spinors, we can define the density matrix as 
\begin{equation}
\rho_{\alpha\beta} \left( \mathbf{x},\mathbf{y},t\right)= \langle \mathbf{x}%
,\alpha| \hat \rho | \mathbf{y},\beta \rangle = \sum_i p_i \psi_{i}\left( 
\mathbf{x},\alpha,t\right)\psi_{i}^\dag\left(\mathbf{y},\beta,t\right)\, ,
\end{equation}
where $p_i$ is the probability to have a state ${\psi}_i$. The density
matrix fulfills the von Neumann equation 
\begin{equation}  \label{vonneumann}
i\hbar\frac{\partial\hat\rho}{\partial t}=\left[\hat H,\hat\rho\right]\, .
\end{equation}

Once the density matrix has been defined, we can define the
Wigner-Stratonovich transform \cite{strato} as 
\begin{equation}
W_{\alpha \beta }(\mathbf{x},\mathbf{p},t)=\int \frac{d^{3}z}{(2\pi \hbar
)^{3}}\exp \left[ -\frac{i}{\hbar }\mathbf{z}\cdot \Phi \right] \rho
_{\alpha \beta }\left( \mathbf{x}+\frac{\mathbf{z}}{2};\mathbf{x}-\frac{%
\mathbf{z}}{2},t\right) \,,  \label{wstrans}
\end{equation}%
where the phase 
\begin{equation}
\Phi =\mathbf{p}-\frac{q}{c}\int_{-1/2}^{1/2}d\eta \mathbf{A}\left( \mathbf{x%
}+\eta \mathbf{z},t\right) \,
\end{equation}%
is used to ensure gauge invariance of the resulting distribution function.
The Wigner-Stratonovich transform has the property that it must to be taken
separately for each component of the 2-by-2 density matrix.

Different approaches to construct a kinetic theory from the
Wigner-Stratonovich transformation are discussed in Ref.\ \cite{zamanian}.
Following this reference we here define a scalar distribution function $f(%
\mathbf{x},\mathbf{p},\mathbf{s},t)$ in the extended phase-space \cite%
{Phase-space-Note} where $\mathbf{s}$ is a vector of unit length. This
distribution function satisfies that 
\begin{equation}
f(\mathbf{x},\mathbf{s},t)=\int d^{3}pf(\mathbf{x},\mathbf{p},\mathbf{s}%
,t)\,,
\end{equation}%
gives the probability to find the particle at position $\mathbf{x}$ with
spin-up in the direction of $\mathbf{s}$, and 
\begin{equation}
f(\mathbf{p},\mathbf{s},t)=\int d^{3}xf(\mathbf{x},\mathbf{p},\mathbf{s}%
,t)\,,
\end{equation}%
gives the probability to find the particle with momentum $\mathbf{p}$ with
spin-up in the direction of $\mathbf{s}$. Using the Wigner-Stratonovich
transformation, the scalar distribution function will be defined as \cite%
{zamanian} 
\begin{align}
f(\mathbf{x},\mathbf{p},\mathbf{s},t)& =\frac{1}{4\pi }\sum_{\alpha ,\beta
=1}^{2}\left( 1+\mathbf{s}\cdot \bm\sigma \right) _{\alpha \beta }W_{\beta
\alpha }(\mathbf{x},\mathbf{p},t)  \notag \\
& =\frac{1}{4\pi }\mathrm{tr}(1+\mathbf{s}\cdot \bm\sigma )W(\mathbf{x},%
\mathbf{p},t),  \label{spintrans}
\end{align}%
where $\mathrm{tr}$ denotes the trace over the spin indices. We recall that
the expectation value polarization density is now given by 
\begin{equation}
\left\langle \bm\sigma \right\rangle (\mathbf{x},t)=\mathrm{tr}[\bm\sigma
\rho (\mathbf{x},\mathbf{y},t)]=3\int d^{3}pd^{2}sf(\mathbf{x},\mathbf{p},%
\mathbf{s},t)\mathbf{s},  \label{spint-calc}
\end{equation}%
where we stress the need for the factor 3. This follows from the form of the
transformation \eqref{spintrans} and is needed to compensate for the quantum
mechanical smearing of the distribution function in spin space. Furthermore,
it should be stressed that the independent spin variable $\mathbf{s}$
constructed in (\ref{spintrans}) generates the rest frame expression for the
spin. In our theory, which is only weakly relativistic, this has limited
consequences. The relation between the rest frame spin $\mathbf{s}$, and the
spatial part of a the spin four-vector $\mathbf{S}$ is given by $\mathbf{S=s}%
+[\gamma ^{2}/(\gamma +1)](\mathbf{v\cdot s)v/}c^{2}$ \cite{Jackson}, where
the kinematic quantities (i.e. the gamma factor $\gamma $ and the velocity $%
\mathbf{v}$) can be expressed in terms of $\mathbf{p}$ and $\mathbf{s}$ (see
below). Since our weakly relativistic theory presented here is only
concerned with spin-dependent terms up to order $\mathbf{v/}c$, the
difference between $\mathbf{S}$ and $\mathbf{s}$ may be overlooked for the
most part, e.g. when computing the magnetization current density.

\section{Evolution equation for the scalar distribution function}

Using the above formalism we obtain a fully gauge invariant Vlasov-like
evolution equation for charged particles. One of the most basic quantum
effects is the tendency for the wave function to spread out. In the
non-relativistic version of the theory \cite{zamanian} this effect end up in
operators $\sin (\hbar \nabla _{x}\cdot \nabla _{p})$ and $\cos (\hbar
\nabla _{x}\cdot \nabla _{p})$ acting on the fields and the distribution
function, where the operators can be defined through the trigonometric
Taylor-expansions \cite{taylorexp-note}. In our present theory we will view
the spatial gradient operator $\nabla _{x}$ as a small parameter, and drop
terms of order $\nabla _{x}^{3}$ or smaller, which means dropping particle
dispersive effects, that are smaller by a factor of the order $\delta ^{2}$
where $\delta $ is the characteristic de Broglie wavelength over the
macroscopic scale length. The other approximation made is to only account
for weakly relativistic effects, as described above. This implies that only
terms up to first order in the velocity is kept, and that the gamma factor
is put to unity. The evolution equation is found using the transformations %
\eqref{wstrans} and \eqref{spintrans} on the evolution equation %
\eqref{vonneumann}, which together with the above given approximations
results in 
\begin{align}
0& =\frac{\partial f}{\partial t}+\left\{ \frac{\mathbf{p}}{m}+\frac{\mu }{%
2mc}\mathbf{E}\times \left( \mathbf{s}+\nabla _{s}\right) \right\} \cdot
\nabla _{x}f  \notag \\
& +q\left( \mathbf{E}+\frac{1}{c}\left\{ \frac{\mathbf{p}}{m}+\frac{\mu }{2mc%
}\mathbf{E}\times \left( \mathbf{s}+\nabla _{s}\right) \right\} \times 
\mathbf{B}\right) \cdot \nabla _{p}f  \notag \\
& +\frac{2\mu }{\hbar }\mathbf{s}\times \left( \mathbf{B}-\frac{\mathbf{p}%
\times \mathbf{E}}{2mc}\right) \cdot \nabla _{s}f  \notag \\
& +\mu \left( \mathbf{s}+\nabla _{s}\right) \cdot \partial _{x}^{i}\left( 
\mathbf{B}-\frac{\mathbf{p}\times \mathbf{E}}{2mc}\right) \partial _{p}^{i}f-%
\frac{\hbar ^{2}q}{8m^{2}c^{2}}\partial _{x}^{i}(\nabla \cdot \mathbf{E}%
)\partial _{p}^{i}f\,,  \label{vlasovc}
\end{align}%
where $\mathbf{p}$ is the momentum (which is related to the velocity through
the spin; see below) and $\mu =\hbar q/2mc$ (or $\mu =g\hbar q/4mc$).

The evolution equation \eqref{vlasovc} has three new effects compared to the
equation in Ref.\ \cite{zamanian} for spin-$1/2$ particles. The first one is
the Thomas precession effect where the previous theory \cite{zamanian} is
extended by the substitution $\mathbf{B}\rightarrow \mathbf{B}-\mathbf{p}%
\times \mathbf{E}/{2mc}$ in the fourth and fifth terms of Eq. \eqref{vlasovc}
. This effect comes from the spin-orbit coupling contribution in Hamiltonian %
\eqref{eq:pos} and, therefore, it is directly coupled with the evolution of
the spin. The second new effect is the last term which is associated to the
Darwin term. This term introduces the Zitterbewegung effect of the electron,
and is the only contribution proportional to $\hbar ^{2}$. The third effect
is seen in the velocity-momentum relation, which is highlighted in the
second and third terms. In Eq.~\eqref{vlasovc} the term in $\{\}$ brackets
resembles a velocity which has been modified by the spin, which will be
discussed in some detail below. Finally, we point out that the factor in
front of $\nabla _{s}f$ in the third term is indeed given by $d\mathbf{s/}dt$
\cite{Jackson}, i.e. the \textit{laboratory rate of change} of the \textit{%
rest frame value} of the spin. Thus we note that Eq. (\ref{vlasovc}) is
consistent with the interpretation of $\mathbf{s}$ as the rest frame
variable for the spin.

Next we consider the relation between the velocity and momentum. In order to
relate this variables we use the Heisenberg equation of motion for the
velocity operator%
\begin{equation}
\hat{\mathbf{v}}=\frac{1}{i\hbar }[\hat{\mathbf{x}},\hat{H}].
\end{equation}%
For the Hamiltonian \eqref{eq:pos} we then get 
\begin{equation}
\hat{\mathbf{v}}=\frac{1}{m}\left( -i\hbar \nabla -\frac{q}{c}\mathbf{A}%
\right) -\frac{\mu }{2mc}\bm{\sigma}\times \mathbf{E},
\end{equation}%
where we have neglected the last term in the Hamiltonian to simplify the
equation slightly, (see the discussion about the mass correction below). We
now recall that the Wigner transformation for an operator is multiplied by a
factor $(2\pi \hbar )^{3}$, as compared to the the Wigner transformation for
the density matrix. Similarly, for the spin transformation, the
transformation for operators comes with a factor 3. Taking this into account
and calculating the Wigner and Q transformation of the operator above gives
the final relation 
\begin{equation}
\mathbf{v}=\mathbf{v}(\mathbf{x},\mathbf{p},\mathbf{s},t)=\frac{\mathbf{p}}{m%
}+\frac{3\mu }{2mc}\mathbf{E}\times \mathbf{s}.  \label{velocity}
\end{equation}%
This is the function in extended phase space, which can be used to calculate
the average velocity and the current density of the plasma. An important
question that arises is whether the current density based on the velocity (%
\ref{velocity}) will give the \textit{free current density} or if correspond
to some other physical quantity, e.g. the total current density. This
question is addressed below, where we calculate the energy conservation law
of our system, which confirms that the velocity in Eq. (\ref{velocity}) is
indeed the variable corresponding to the free current density.

In a more general context, the relationship between the momentum and the
velocity is nontrivial. E.g. the spin orbit coupling has been shown to arise
as a Berry phase term \cite{mathur}. For further discussion of this
interesting topic see e.g. \cite{berard, bliokh, bliokh2, gosselin}.

When obtaining the evolution equation \eqref{vlasovc}, we have not
considered the effect of the mass-velocity correction term in order to get a
more transparent formalism. This term will only produce a correction of the
form $\mathbf{p}/{m}\longrightarrow \mathbf{p}/{m}\,(1+\mathbf{p}%
^{2}/2m^{2}c^{2})$ in the second term. Although this term is of the same
order in an expansion in $1/c$, as compared to other terms that have been
kept, we will not consider it as the classical relativistic terms are
already well-known. Instead we focus on the new effects introduced by the
spin and the Zitterbewegung.

The dynamics of the distribution function given by the Vlasov equation %
\eqref{vlasovc} is in the mean field approximation coupled to the Maxwell
equations in the form 
\begin{equation}
\nabla \cdot \mathbf{E}=4\pi \rho _{T}\,,\qquad \nabla \times \mathbf{B}=%
\frac{\partial \mathbf{E}}{\partial t}+4\pi \mathbf{J}_{T}\,,
\end{equation}%
where the total charge density and total current density are given by 
\begin{equation}
\rho _{T}=\rho _{F}+\nabla \cdot \mathbf{P}\,\qquad \mathbf{J}_{T}=\mathbf{J}%
_{F}+\nabla \times \mathbf{M}+\frac{\partial \mathbf{P}}{\partial t}\,.
\label{Tcurrent}
\end{equation}%
In the above expressions, the free charge density is 
\begin{equation}
\rho _{F}=q\int d\Omega f\,,  \label{charge}
\end{equation}%
where $d\Omega =d^{3}vd^{2}s$ is the integration measure performed over the
three velocity variables and the two spin degrees of freedom. The spin
vector has a fixed unity length and it is thus convenient to use spherical
coordinates $(\varphi _{s},\theta _{s})$ do describe it. The free current
density is given by 
\begin{equation}
\mathbf{J}_{F}=q\int d\Omega \left( \frac{\mathbf{p}}{m}+\frac{3\mu }{2mc}\,%
\mathbf{E}\times \mathbf{s}\right) f.  \label{Free-current}
\end{equation}%
With these charge and current densities, the conservation of charge is
obtained as from \eqref{vlasovc} to be $\partial _{t}\rho _{F}+\nabla \cdot 
\mathbf{J}_{F}=0$. Furthermore the magnetization $\mathbf{M}$ and the
polarization $\mathbf{P}$ are both due to the spin and they are calculated
respectively as 
\begin{equation}
\mathbf{M}=3\mu \int d\Omega \mathbf{s}f  \label{Magnetization}
\end{equation}%
and 
\begin{equation}
\mathbf{P}=-3\mu \int d\Omega \frac{\mathbf{s}\times \mathbf{p}}{2mc}f.
\label{Polarization}
\end{equation}

The system of Maxwell's equations with the magnetization (\ref{Magnetization}%
) and polarization (\ref{Polarization}) and free current density (\ref%
{Free-current}), together with our main equation (\ref{vlasovc}), satisfies
an energy conservation law of the form 
\begin{equation}
\partial _{t}W+\nabla \cdot \mathbf{K}=0.  \label{Energy-conservation}
\end{equation}%
Here the total energy density $W$ is given by 
\begin{equation}
W=\frac{1}{2}\left( |\mathbf{E}|^{2}+|\mathbf{B}|^{2}\right) +\int d\Omega
\left( \frac{\mathbf{p}^{2}}{2m}-3\mu \mathbf{s}\cdot \mathbf{B}\right) f\,,
\label{energy-density}
\end{equation}%
and the energy flux vector $\mathbf{K}$ is given by 
\begin{equation}
\mathbf{K}=\mathbf{E}\times \left( \mathbf{B}-\mathbf{M}\right) +\int
d\Omega \left( \frac{\mathbf{p}^{2}}{2m}+3\mu \left( \mathbf{B}-\frac{%
\mathbf{p}\times \mathbf{E}}{2mc}\right) \cdot \mathbf{s}\right) \mathbf{v}%
f\,.  \label{energy-flux}
\end{equation}%
Apparently the first terms in (\ref{energy-density}) constitute the
electromagnetic field energy density, and the integral term is the combined
kinetic and magnetic dipole energy density. The first term of (\ref%
{energy-flux}) is the Poynting vector, whereas the latter represents the
combined flux of kinetic energy density and magnetic dipole energy. This
energy conservation equation is a generalization of previous results for
semi-classical theories for spin-1/2 plasmas \cite{brodin}. It should be
noted that although the theory presented here contains approximation, e.g.
due to the weakly relativistic assumptions, the conservation law (\ref%
{Energy-conservation}) is an exact property of the presented model.

\section{Linearized theory}

In the present section, we are going to study the influence of the
spin-orbit coupling and of the Darwin term on linear wave propagation. For
this purpose we linearize the evolution equation \eqref{vlasovc}, where the
variables are separated into equilibrium and perturbed quantities (using the
subindices $0$ and $1$, respectively, to denote them). Thus, the
distribution function will be $f=f_{0}+f_{1}$, and the electric and magnetic
field could be written as $\mathbf{E}=\mathbf{E}_{1}$ and $\mathbf{B}=%
\mathbf{B}_{0}+\mathbf{B}_{1}$ respectively. The evolution equation to
linear order becomes 
\begin{widetext}
 \begin{align}
  \frac{\partial f_1}{\partial t}&+\frac{\mathbf p}{m}\cdot\nabla_x f_1+\frac{q}{mc}\mathbf p\times\mathbf B_0\cdot\nabla_p f_1+\frac{2\mu}{\hbar}\mathbf s\times\mathbf B_0\cdot\nabla_s f_1+\mu \nabla_{xi}\left[\mathbf B_0\cdot(\mathbf s+\nabla_s)\right]\nabla_{pi}f_1\nonumber\\
&=-q\mathbf E_1\cdot\nabla_p f_0-\frac{\mu}{2mc}\mathbf E_1\times\left(\mathbf s+\nabla_s\right)\cdot\nabla_xf_0-\frac{q}{mc}\mathbf p\times\mathbf B_1\cdot\nabla_p f_0-\mu\nabla_{xi}\left[\mathbf B_1\cdot(\mathbf s+\nabla_s)\right]\nabla_{pi}f_0\nonumber\\
& +\frac{\mu}{2mc}\nabla_{xi}\left[(\mathbf p\times\mathbf E_1)\cdot(\mathbf s+\nabla_s)\right]\nabla_{pi}f_0-\frac{q\mu}{2mc^2}\left[\mathbf E_1\times(\mathbf s+\nabla_s)\right]\times\mathbf B_0\cdot\nabla_p f_0-\frac{2\mu}{\hbar}\mathbf s\times\mathbf B_1\cdot\nabla_s f_0\nonumber\\
&+\frac{\mu}{\hbar mc}\mathbf s\times(\mathbf p\times\mathbf E_1)\cdot\nabla_s f_0+\frac{\hbar^2 q}{8m^2c^2}\nabla_{xi}(\nabla\cdot\mathbf E_1)\nabla_{pi}f_0\, .
\label{vlasovclinear} \end{align}
\end{widetext}

In the following, we only study electrostatic modes (e.g. $\mathbf{B}_{1}=0$
in \eqref{vlasovclinear}) propagating along $\mathbf{B}_{0}$, as this gives
a good illustration of the contribution from the relativistic terms, that
are due to the Zitterbewegung effect and the spin-orbit coupling.

\subsection{Darwin term contribution}

Firstly we want to focus on the effect associated with Zitterbewegung. The
Zitterbewegung is a rapid oscillatory motion of the electron which implies
that if an instantaneous measurement of its velocity is performed, the
result is the speed of light. The amplitude of the oscillatory motion is $%
x_{osc}\sim \hbar /2mc$ \cite{strange}, which means that the electron cannot
be localized, but is rapidly oscillating in a volume of the order of the
cube of the Compton wavelength. The Zitterbewegung is a quantum relativistic
effect and it is related to particle-antiparticle nature of the Dirac theory
and to the nature of the spin. At the present, there is a growing interest
in the detection of effects as the Zitterbewegung \cite{gerri}.

The effect of the Zitterbewegung of the electron is introduced in the last
term of Eq. \eqref{vlasovclinear}, the Darwin contribution, which represents
the smeared out electrostatic potential field that the electron sees when it
fluctuates over a distance $x_{osc}$.

For the sake of simplicity, we examine the dispersion relation of Langmuir
waves in an unmagnetized plasma. In order to focus on the effects of the
Darwin contribution, we consider a one-dimensional unperturbed momentum
distribution and let $f_{0}\rightarrow $ $f_{0}(p_{z}^{2})\delta
(p_{x})\delta (p_{y})$ (for a more realistic 3D-momentum distribution, even
the electrostatic unmagnetized case couple to the spin terms, as we will see
in the next section). The total distribution function will then have the
form of $f(z,p_{z},t)=f_{0}(p_{z}^{2})+f_{1}(z,p_{z},t)\exp (ikz-i\omega t)$%
, where $f_{0}$ and $f_{1}$ are the equilibrium and the perturbed
distribution functions respectively. Furthermore, we consider a homogeneous
plasma an neglect the motion of the heavy ions. The perturbed electric field
is longitudinal, i.e. $\mathbf{E}_{1}=\hat{z}E_{1}\exp (ikz-i\omega t)$.
Using the evolution equation \eqref{vlasovclinear}, the perturbed
distribution function is then related to the electric field amplitude by 
\begin{equation}
f_{1}=\frac{-iqE_{1}}{\omega -kp_{z}/m}\left( 1+\frac{\hbar ^{2}k^{2}}{%
8m^{2}c^{2}}\right) \frac{\partial f_{0}}{\partial p_{z}}\,.
\label{f1darwinikE}
\end{equation}%
Combining Eq. (\ref{f1darwinikE}) with Poisson equation $\nabla \cdot 
\mathbf{E}_{1}=ikE_{1}=4\pi q\int d\Omega f_{1}$, we obtain the dispersion
relation 
\begin{equation}
1=\frac{\omega _{p}^{2}}{k^{2}}\left( 1+\frac{\hbar ^{2}k^{2}}{8m^{2}c^{2}}%
\right) \int_{-\infty }^{\infty }dp_{z}\frac{\widehat{f_{0}}}{\left(
p_{z}/m-\omega /k\right) ^{2}}\,  \label{integf0}
\end{equation}%
where the re-normalized distribution function $\widehat{f_{0}}$ fulfills$%
\int_{-\infty }^{\infty }dp_{z}$ $\widehat{f_{0}}=1$. \ For phase velocities
larger than than the characteristic spread in $p_{z}/m$, we can Taylor
expand the denominator, and write the dispersion relation as

\begin{equation}
\omega ^{2}=\omega _{p}^{2}\left( 1+\frac{\hbar ^{2}k^{2}}{8m^{2}c^{2}}%
\right) \left( 1+\frac{k^{2}\left\langle p_{z}^{2}\right\rangle }{%
m^{2}\omega ^{2}}\right)
\end{equation}%
where $\left\langle p_{z}^{2}\right\rangle =\int_{-\infty }^{\infty
}dp_{z}p_{z}^{2}\widehat{f_{0}}$ is the average of the squared momentum.
Here the Landau damping term has been dropped, since the resonance is
assumed to lie in the tail of the distribution.

\begin{figure}[t]
\begin{center}
\includegraphics[scale = 0.95]{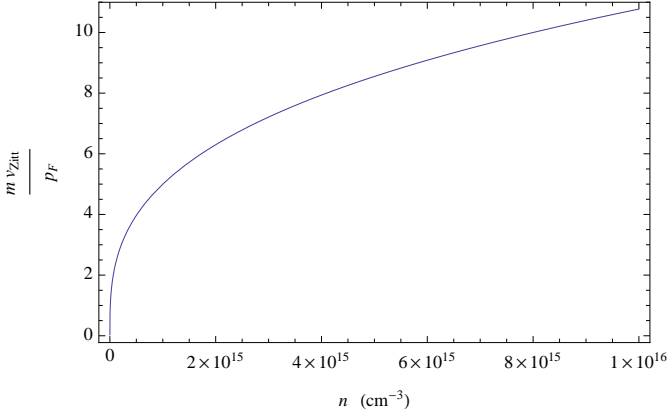}
\end{center}
\caption{The parameter $m v_{\text{\rm Zitt}} / p_F$, relevant as a comparison
when $T \rightarrow 0$, where $p_F$ is the Fermi momentum, is plotted as a
function of density. This illustrates the relative importance of the
Zitterbewegung and the Fermi momentum, i.e. the Fermi pressure, in the
dispersion relation (\protect\ref{31}) above. }
\label{fig:zitt}
\end{figure}

The term proportional to $\hbar ^{2}k^{2}$ is the Zitterbewegung
contribution to electrostatic modes. For not too large wave-numbers it is
more significant than quantum contributions from the Bohm potential \cite%
{Manfredi} that scales as $\hbar ^{2}k^{4}$. Provided the wave-numbers are
small, $\omega \simeq \omega _{p}$, and the dispersion relation can be
further approximated as 
\begin{equation}
\omega^2 = \omega_p^2 + k^2\left( \frac{\left\langle
p_z^2\right\rangle }{m^2} + \frac{v_{\rm Zitt}^2}{2} \right) \, ,  
\label{31}
\end{equation}%
where $v_{\rm Zitt} = \hbar \omega_p/2mc$ can be understood as a velocity
response of the plasma $v_{\rm Zitt}\sim x_{\rm osc}\omega_p$ to the rapid
oscillations of the Zitterbewegung motion. The term $v_{\rm Zitt}^2k^2$
comes from the fact that the electron sees a smeared out electrostatic
potential, and therefore, a gradient of the electric field and a force of
the order $\hbar^2\nabla (\nabla \cdot \mathbf{E})/m^2c^2$. Similar
to the effects of the Fermi pressure (which gives a non-zero $\left\langle
p_{z}^{2}\right\rangle $ even when the temperature goes to zero) this
implies a non-zero group velocity of the electron plasma waves even in the
cold case (see also Fig. 2 for a comparison between the Fermi statistics and
the Zitterbewegung, relevant for Eq. (\ref{31}) when $T\rightarrow 0$). We
note that the effect of the Darwin term related to Zitterbewegung becomes
important for high-density plasmas when $\hbar \omega_p/mc^2$ is not
too much smaller than unity.

\subsection{Spin-orbit coupling contribution}

The spin-orbit effect of Hamiltonian \eqref{eq:pos} appears as the Thomas
precession correction (see e.g. Ref. \cite{Jackson-2}) of the magnetic field
in the fourth and fifth terms of Eq. \eqref{vlasovc}.

Although this contribution can introduce interesting corrections to
different types of wave modes, in this work we are going to follow the
previous spirit and we analyze the quantum corrections to Langmuir waves,
this time in a magnetized plasma. These modes have been studied previously
in Ref.\ \cite{Petter} in a phenomenological relativistic formalism where
the appropriate Thomas precession factor of $1/2$ was not used. We consider
again longitudinal electrostatic modes $\mathbf{E}_{1}=\hat{z}\,E_{1}\exp
(ikz-i\omega t)$, but now propagating along an external magnetic field $%
\mathbf{B}_{0}=B_{0}\hat{z}$. As will be seen below, this will give an
illustrative example of how the spin-orbit coupling modifies the usual
dispersion relation. The distribution function will be taken to be of the
form $f(z,\mathbf{p},\mathbf{s},t)=f_{0}(\mathbf{p}^{2},\theta _{s})+f_{1}(z,%
\mathbf{p},\mathbf{s},t)\exp (ikz-i\omega t)$. We use cylindrical
coordinates for the momentum, i.e. $\mathbf{p}=\hat{x}p_{\perp }\cos \varphi
_{p}+\hat{y}\mathbf{p}_{\perp }\sin \varphi _{\perp }+\hat{z}p_{z}$ with $%
\mathbf{p}^{2}=p_{\perp }^{2}+p_{z}^{2}$. Furthermore, we use spherical
coordinates for the spin, i.e. $\mathbf{s}=\hat{x}\sin \theta _{s}\cos
\varphi _{s}+\hat{y}\sin \theta _{s}\sin \varphi _{s}+\hat{z}\cos \theta
_{s} $.

At first order, taking the Fourier analysis of the evolution equation %
\eqref{vlasovclinear}, we have 
\begin{widetext}
\begin{align}
\left(\frac{\partial}{\partial t}+\frac{\mathbf p}{m}\cdot\nabla_x-\omega_c\frac{\partial}{\partial\varphi_p}-\omega_{cg}\frac{\partial}{\partial\varphi_s}\right)f_1&=-qE_1\left(1+\frac{\hbar^2k^2}{8m^2 c^2}\right)\frac{\partial f_0}{\partial p_z}\nonumber\\
&-\frac{ik\mu p_\perp E_1}{2mc}\left(\sin\theta_s+\cos\theta_s\frac{\partial}{\partial\theta_s}\right)\left(\cos\varphi_p\sin\varphi_s-\sin\varphi_p\cos\varphi_s\right)\frac{\partial f_0}{\partial p_z}\nonumber\\
&+\frac{\mu}{\hbar mc}p_\perp E_1\left(\cos\varphi_p\cos\varphi_s+\sin\varphi_p\sin\varphi_s\right)\frac{\partial f_0}{\partial \theta_s}\nonumber\\
&-\frac{q\mu B_0 E_1}{2mc^2}\left(\sin\theta_s\cos\varphi_s+\cos\theta_s\cos\varphi_s\frac{\partial}{\partial \theta_s}\right)\left(\cos\varphi_p\frac{\partial f_0}{\partial p_\perp}\right)\nonumber\\
&-\frac{q\mu B_0 E_1}{2mc^2}\left(\sin\theta_s\sin\varphi_s+\cos\theta_s\sin\varphi_s\frac{\partial}{\partial \theta_s}\right)\left(\sin\varphi_p\frac{\partial f_0}{\partial p_\perp}\right)
\label{spinO1}\end{align}
\end{widetext}where $\omega _{c}=qB_{0}/mc$ is the cyclotron frequency and $%
\omega _{cg}=(g/2)\omega _{c}$ is the spin precession frequency \cite{brodin}%
. We note that the perturbed distribution function can be solved for in
terms of the orthogonal eigenfunctions $\psi _{n}$ to the right hand side
operator \cite{brodin,Petter}. Accordingly we make the expansion 
\begin{equation}
f_{1}=\frac{1}{\sqrt{2\pi }}\sum_{n=-\infty }^{\infty }\sum_{n^{\prime
}=-\infty }^{\infty }g_{n,n^{\prime }}(p_{\perp },p_{z},\theta _{s})\psi
_{n}(p_{\perp },\varphi _{p})\exp (in^{\prime }\varphi _{s})\,,
\label{basef}
\end{equation}%
where in general 
\begin{eqnarray}
\psi _{n}(p_{\perp },\varphi _{p}) &=&\frac{1}{\sqrt{2\pi }}\exp
[-i(n\varphi _{v}-(k_{\perp }p_{\perp }/\omega _{c}m)\sin \varphi _{p})] \\
&=&\frac{1}{\sqrt{2\pi }}\sum_{n^{\prime \prime }=-\infty }^{\infty
}J_{n^{\prime \prime }}\left( \frac{k_{\perp }p_{\perp }}{m\omega _{c}}%
\right) e^{i(n-n^{\prime \prime })\varphi _{p}}\,,
\end{eqnarray}%
and $J_{n^{\prime \prime }}$ is the Bessel function. However, in this case
for longitudinal mode with $k_{\perp }=0$, then 
\begin{equation}
\psi _{n}(p_{\perp },\varphi _{p})=\frac{1}{\sqrt{2\pi }}e^{in\varphi
_{p}}\,.
\end{equation}%
Using the distribution function \eqref{basef} in Eq. \eqref{spinO1}, and
then multiplying both sides by $\psi _{n}^{\ast }e^{-im\varphi _{s}}$ and
integrating over $\varphi _{p}$ and $\varphi _{s}$, we find that the only
terms that survive in the sum \eqref{basef} are $g_{0,0}$ and $g_{\pm 1,\mp
1}$ \cite{zamanian,brodin,Petter}. Thus, we find the solution for $f_{1}$ as 
\begin{widetext}
\begin{align}
 f_1&=\left(\frac{-iqE_1(1+{\hbar^2k^2}/{8m^2c^2})}{\omega-kp_z/m}\right)\frac{\partial f_0}{\partial p_z}+\frac{i\mu p_\perp E_1}{2\hbar mc}\left[\frac{e^{i(\varphi_p-\varphi_s)}}{\omega-\Delta\omega_c-kp_z/m}+\frac{e^{i(\varphi_s-\varphi_p)}}{\omega+\Delta\omega_c-kp_z/m}\right] \frac{\partial f_0}{\partial \theta_s}\nonumber\\
&+\frac{ik\mu p_\perp E_1}{4mc}\left[\frac{e^{i(\varphi_p-\varphi_s)}}{\omega-\Delta\omega_c-kp_z/m}-\frac{e^{i(\varphi_s-\varphi_p)}}{\omega+\Delta\omega_c-kp_z/m}\right]\left(\sin\theta_s+\cos\theta_s\frac{\partial}{\partial\theta_s}\right) \frac{\partial f_0}{\partial \theta_s}\, ,
\label{spinO2}
\end{align}
\end{widetext}where $\Delta \omega _{c}=\omega _{cg}-\omega _{c}$. 
The expression \eqref{spinO2} is combined with 
\begin{equation}
-i\omega E_{1}=-4\pi J_{Tz}\,,  \label{Max1}
\end{equation}%
is used to deduce the dispersion relation. Due to the dependence on the
angles $\varphi _{s}$ and $\varphi _{p}$, the first term of \eqref{spinO2}
give raise to a free current density, whereas the other terms give raise to
a polarization current density. The magnetization current density vanish
identically. Combining \eqref{spinO2} and \eqref{Max1} we find the
dispersion relation 
\begin{widetext}
\begin{align}
\omega&=-\left(\omega_p^2+\frac{1}{2}v_{\rm Zitt}^2k^2\right)\int d\Omega p_z\frac{\frac{\partial \widehat{f_{0}}}{\partial p_z}}{\omega-kp_z/m}+\frac{6\pi^3\mu^2\omega}{\hbar m^2 c^2}\int d\Omega p_\perp^2\frac{\widehat{f_{0}}}{\partial \theta_s}\sin\theta_s\left(\frac{1}{\omega-\Delta\omega_c-kp_z/m}-\frac{1}{\omega+\Delta\omega_c-kp_z/m}\right)\nonumber\\
&+\frac{3\mu^2\omega k\pi^3}{m^2c^2}\int d\Omega p_\perp^2\left(\sin\theta_s+\cos\theta_s\frac{\partial}{\partial\theta_s}\right)\frac{\partial \widehat{f_{0}}}{\partial p_z}\left(\frac{1}{\omega-\Delta\omega_c-kp_z/m}+\frac{1}{\omega+\Delta\omega_c-kp_z/m}\right)\, ,
\label{dipsspinorbit}\end{align}
\end{widetext}which is general for $\widehat{f_{0}}$ (where we have again
used the distribution function re-normalized as $\int \widehat{f_{0}}d\Omega
=1$).

As an example, let us examine an equilibrium distribution function with the
form of a Maxwellian distribution and a spin dependent part \cite{zamanian} 
\begin{align}
\widehat{f_{0}}(\mathbf{p}^2,\theta_s) = \frac{1}{N_M}e^{-\mathbf{p}^2/m^2
v_t^2} & \left[e^{\mu B_0/k_B T}(1+\cos\theta_s) \right.  \notag \\
& \left. +e^{-\mu B_0/k_B T}(1-\cos\theta_s)\right]\, ,  \label{distrf0spin}
\end{align}
where $T$ is the temperature and $k_B$ is the Boltzmann constant, and the
normalization factor is $N_M=4\pi(\pi m^2 v_t^2)^{3/2}\cosh (\mu B_0/k_BT)$.
We note that the expression \eqref{distrf0spin} is the thermodynamic
equilibrium distribution for a plasma of moderate density where the
magnitude of the chemical potential is large \cite{distribution-note}.

To simplify the integrals we will consider the frequency range where the
wave frequency $\omega $ be close to resonance with $\Delta \omega _{c}$. In
this case, we neglect in \eqref{dipsspinorbit} the terms with the
denominators $1/(\omega +\Delta \omega _{c}-p_{z}k/m)$ because they are
small compared with the terms with denominators $1/(\omega -\Delta \omega
_{c}-p_{z}k/m)$. We are also going to take the limit when $\omega -\Delta
\omega _{c}\gg p_{z}k/m$. Thus, using the equilibrium distribution function %
\eqref{distrf0spin}, Taylor expanding the denominators, neglecting the poles
in $\omega =p_{z}k/m$ and $\omega -\Delta \omega _{c}=p_{z}k/m$, and
integrating over $d\Omega $, we finally find the dispersion relation for
Langmuir waves with spin-orbit coupling and Darwin effects 
\begin{widetext}
\begin{align}
& \omega^2\left\{1+\frac{\hbar^2\pi^2\omega_p^2}{8m^2c^4}\left[\frac{k^2 v_t^2}{(\omega-\Delta\omega_c)^2}+\frac{3k^4v_t^4}{2(\omega-\Delta\omega_c)^4}\right]+\frac{\hbar\pi^2\omega_p^2v_t^2}{4mc^4}\tanh\left(\frac{\mu B_0}{k_BT}\right)\left[\frac{1}{\omega-\Delta\omega_c}+\frac{k^2v_t^2}{2(\omega-\Delta\omega_c)^3}\right]\right\}\nonumber\\
&=\left(\omega_p^2+\frac{1}{2}v_{\rm Zitt}^2k^2\right)\left(1+\frac{3k^2v_t^2}{2\omega^2}\right)\, . 
\end{align}

\end{widetext}The coefficient in front of the terms with denominators $%
\propto \omega -\Delta \omega _{c}$ are typically small, except for very
strong magnetic fields. Thus excluding the regime of an extremely strong
external field, the frequency of the spin modes will be close to resonance,
i.e. fulfill $\omega \simeq \Delta \omega _{c}$. More specifically, the
deviation from exact resonance is of the order $\left( \omega -\Delta \omega
_{c}\right) /\Delta \omega _{c}\sim (\hbar \Delta \omega _{c}/mc^{2})\tanh
(\mu B_{0}/k_{B}T)$. We note that spin induced modes with $\omega \simeq
\Delta \omega _{c}$ have already been found by Ref. \cite{brodin} without
the inclusion of spin-orbit coupling. However, it should be noted that the
present wave mode is quite different from that previously found. In
particular, the field is now completely electrostatic (whereas it was found
to be completely electromagnetic in the previous case), and the present wave
mode exists in the long wavelength regime, whereas the former wave mode \cite%
{brodin} was dependent on a short wavelength, i.e. of the order of the
electron gyroradius or shorter. Finally we note that even in the absence of
an external magnetic field, a finite contribution from the electron spin
remains, together with the Darwin contribution. In the absence of resonances
we note that both these quantum contributions require a very high plasma
density to be significant.

\section{Conclusions}

The starting point for our theory is the FW-transformation, applied on the
Dirac equation, which is used to pick out the positive energy states. Using
the Wigner-Stratonovich transformation (\ref{wstrans}) on the density
matrix, together with a Q-transformation \cite{zamanian} a scalar
distribution function (\ref{spintrans}) can be defined. In the weakly
relativistic limit, the evolution equation is given by (\ref{vlasovc}) to
leading order in the expansion parameters $\hbar L^{-1}/p$ and $\mu B/mc^{2}$%
, where $L$ is a characteristic macroscopic scale length, $p$ a
characteristic momentum of particles, and $B$ a characteristic magnitude of
the magnetic field. In addition to the magnetic dipole force and spin
precession \cite{zamanian}, which is included already in the Pauli
Hamiltonian, Eq.\ (\ref{vlasovc})) also contains spin-orbit interaction,
including the Thomas factor, and also the contribution from the Darwin term.
A complication in the (weakly) relativistic theory that should be noted is
the relation between momentum and velocity (\ref{velocity}), which now is
dependent on the spin variable. Moreover, when closing the system, the
polarization (\ref{Polarization}) associated with the spin must be included
in the current and charge density (\ref{Tcurrent}), and the expression for
the free current density (\ref{Free-current}) is affected due to the
aforementioned relation (\ref{velocity}). It should also be stressed that
the spin variable used to define $f(\mathbf{x},\mathbf{p},\mathbf{s},t)$
refers to the rest-frame spin, which is convenient, as two spherical angles
for the spin variables is sufficient.

In order to illustrate the theory, we have presented examples of
electrostatic interaction in magnetized and non-magnetized plasmas. By
picking a 1-D unperturbed distribution function, the influence of the Darwin
term, which is associated with Zitterbewegung, is highlighted in the case of
a non-magnetized plasma. For a magnetized plasma, the spin-orbit terms leads
to new types of resonances for electrostatic waves, which involve the
combined effect of orbital and spin-precession motion.

The theory presented here is of most interest for systems where at least one
of the parameters $\hbar \omega _{p}/mc^{2}$ or $\mu B/mc^{2}$ are not too
small. Examples include e.g. laser-plasma interactions, astrophysical
objects (e.g. white dwarf stars), solid state plasmas and strongly
magnetized systems. The present paper is a first step to reach a fully
relativistic quantum theory of plasmas.


\acknowledgments F. A. A. is very grateful to the hospitality of Ume{\aa }
University, and he thanks M. Estrella for support. This work is supported by
the Baltic Foundations, the European Research Council under Contract No.\
204059-QPQV, and the Swedish Research Council under Contract Nos.\ 2007-4422
and 2010-3727.


\begin{thebibliography}{99}
\bibitem{Phase-space-Note} Note that the extended phase space of our theory 
\cite{zamanian} is not itself a phase space, since no angular position
variable is present. Instead it is a regular phase space $(\mathbf{x},%
\mathbf{p})$ plus a spin variable.
\bibitem{taylorexp-note} More generally, the final equation can be written
as an integro-differential equation.
\bibitem{distribution-note} More generally, for a plasma with high density
where degeneracy effects is important, the expression (\ref{distrf0spin})
can be replaced with a sum of two Fermi-Dirac type of distributions, where
the Fermi energy is substituted according to $E_F\rightarrow E_F+\mu B_0$
for the spin-up distribution, and $E_F\rightarrow E_F-\mu B_0$ for the
spin-down species. The general thermodynamic equilibrium distribution in a
homogeneous magnetized plasma also accounting for Landau quantization can be
found in Ref. \cite{zamanian}.
\end{thebibliography}
\end{document}